\documentclass[twocolumn,showpacs,prb,preprintnumbers,amsmath,amssymb]{revtex4}
\usepackage{graphicx}
\usepackage{dcolumn}
\usepackage{multirow}
\usepackage{epsfig}
\begin{document}
\title{Ab initio lattice dynamics simulations and inelastic neutron scattering spectra for studying phonons in BaFe$_{2}$As$_{2}$: Effect of structural phase transition, structural relaxation and magnetic ordering}
\author{M. Zbiri$^{1}$, H. Schober$^1$, M. R. Johnson$^1$, S. Rols$^1$, R. Mittal$^2$, Y. Su$^2$, M. Rotter$^3$, and D. Johrendt$^3$} 
\affiliation{$^1$Institut Max von Laue-Paul Langevin, 6 rue Jules Horowitz, BP 156, 38042 Grenoble Cedex 9, France}
\affiliation{$^2$Juelich Centre for Neutron Science, IFF, Forschunszentrum Juelich, Outstation at FRM II, Lichtenbergstr. 1, D-85747 Garching, Germany}
\affiliation{$^3$Department Chemie und Biochemie, Ludwig-Maximilians-Universitaet Muenchen, Butenandtstrasse 5-13 (Haus D), D-81377 Muenchen, Germany}
\begin{abstract}
We have performed extensive ab initio calculations to investigate phonon dynamics and their possible role in superconductivity in BaFe$_2$As$_2$ and related systems. 
The calculations are compared to inelastic neutron scattering data that offer improved resolution over published data [Mittal et al., PRB {\bf 78} 104514 (2008)], in particular at low frequencies.
Effects of structural phase transition and full/partial structural relaxation, with and without magnetic ordering, on the calculated vibrational density of states are reported. 
Phonons are best reproduced using either the relaxed magnetic structures or the experimental cell.
Several phonon branches are affected by the subtle structural changes associated with the transition from the tetragonal to the orthorhombic phase. 
Effects of phonon induced distortions on the electronic and spin structure have been investigated. 
It is found that for some vibrational modes, there is a significant change of the electronic distribution and spin populations around the Fermi level. 
A peak at 20 meV in the experimental data falls into the pseudo-gap region of the calculation. This was also the case reported in our recent work combined with an empirical parametric calculation [Mittal et al., PRB {\bf 78} 104514 (2008)]. 
The combined evidence for the coupling of electronic and spin degrees of freedom with phonons is relevant to the current interest in  superconductivity in BaFe$_2$As$_2$ and related systems.
\end{abstract}
\maketitle
\section{Introduction}
Every time a new class of superconducting material is discovered another door is opened to an improved understanding of the phenomenon of high 
temperature superconductivity. This holds, in particular, when transition temperatures beyond some 20 K are reached. Examples are the research 
activities triggered by the discovery of superconductivity in doped fullerenes, boron-carbides and MgB$_2$. Despite the undeniable new insights 
procured by these studies, the question of high-temperature superconductivity continues to elude description by physical models. Oxypnictides, a family 
of superconductors with rather high transition temperatures that share structural and electronic two-dimensonality with copper-oxides, offer a new 
class of materials to study high temperature, superconductivity.\\
Superconductivity has been evidenced in iron-oxypnictide LaOFeP~\cite{super1} with T$_c$ $\sim 4$ K, increasing to $\sim 7$ K with F 
doping~\cite{super2}. A similar transition temperature has been measured for nickel-oxypnicitde LaONiP~\cite{super3} (T$_c \sim 4$ K). 
An unexpectedly high-T$_c$ (T$_c \sim 26$ K) was recently discovered in the F-doped iron-oxypnicitde La[O$_{1-x}$F$_x$]FeAs~\cite{super4,super5}. 
The parent compound LaFeAsO offers a unique picture of the chemical bonding; dominantly ionic in the LaO layers and covalent in the FeAs layers. 
Thus charge transfer can be assumed to occur according to (LaO)$^{+}$(FeAs)$^{-}$. It is established that the superconductivity in LaFeAsO originates 
mainly from structural and electronic properties of the FeAs layers. Similar structural and electronic features have been observed in the ternary, 
oxygen-free, iron-arsenide BaFe$_2$As$_2$. The only difference between the two compounds is that the FeAs layers, which are in principle electronically 
identical in both cases, are separated in BaFe$_2$As$_2$ by Ba atoms instead of LaO layers. In this case the charge transfer occurs according to 
Ba$^{2+}_{0.5}$(FeAs)$^{-}$. BaFe$_2$As$_2$ becomes superconducting under pressure at 29 K~\cite{Alireza-2008} and exhibits a phase transition 
around 140 K involving a structural transition (distortion) from tetragonal (high temperature) to orthorhombic (low temperature)~\cite{rotter}. BaFe$_2$As$_2$, 
featuring the simplest possible buffer in between the FeAs-layers, is therefore ideal for studying the coupling of nuclear and electronic degrees 
of freedom.
\begin{figure*}[htbp]
\begin{center}
\epsfig{file=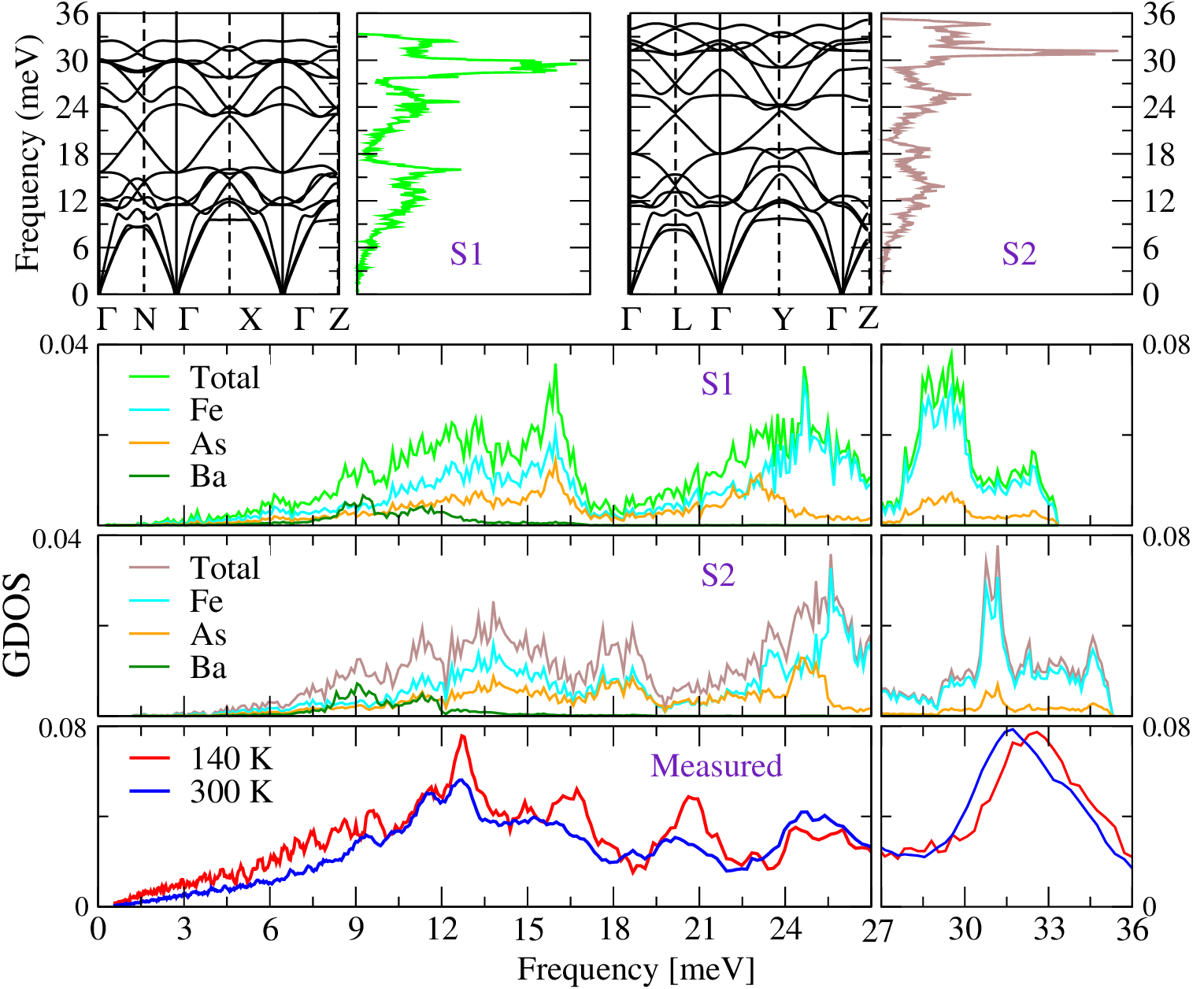, width=15 cm}
\end{center}
\caption{(Color online) Top panels: dispersion relations and vibrational DOS for tetragonal (S1) and orthorhombic (S2) phases. For the tetragonal phase the high symmetry points are N=(1/2,0,0), X=(0,0,1/2) and Z=(1/2,1/2,-1/2). For the orthorhombic phase they are L=(1/2,0,0), Y=(1/2,0,1/2) and Z=(1/2,1/2,0).
Middle panels: Weighted, partial, vibrational DOS and the generalised vibrational DOS for the two phases.
Experimental, generalised vibrational DOS, measured at two temperatures.}
\end{figure*}
\\Phonons and electron-phonon coupling have recently been studied computationally in the parent compound 
LaOFeAs~\cite{phon1,phon2}. It is generally concluded that conventional electron-phonon coupling is insufficient to explain superconductivity in the 
family of Fe-As-based superconductors, while some reports in literature~\cite{Eschrig} indicate the possibility of non-conventional electron-phonon 
coupling. The example of MgB$_2$ showed that phonon coupling in conjunction with particular Fermi surface nesting effects is a viable way to describe 
transition temperatures of up to 40 K~\cite{Pickett-Nature-2002}. 
If phonons do not directly mediate the electron pairing interaction, other excitations are required and spin fluctuations have been considered and 
measured in BaFe$_2$As$_2$~\cite{Boothroyd-PRB}.\\
Experimentally, phonons in Fe-As superconductors have been investigated by inelastic neutron scattering (INS) in fluorine doped LaOFeAs 
\cite{christianson} and BaFe$_2$As$_2$ \cite{mittal}. Thermal neutrons were used, allowing spectra to be measured in neutron down-scattering from 4K up 
to room temperature, but this spectrometer configuration tends to limit energy resolution for low frequency modes, compared to using cold-neutrons in 
an up-scattering configuration. In LaOFeAs, experimental data was compared with published, ab initio phonon calculations \cite{singh}. The agreement 
between experiment and calculation is reasonable, corresponding band frequencies being within a few meV, but this level of inaccuracy leads to strong 
peaks at 30 and 40 meV in the experimental spectra being compared to a single peak at 35 meV in the calculation. In the case of BaFe$_2$As$_2$ the 
experimental resolution was not sufficient to see evidence of the structural phase transition at 140 K in the phonon spectrum. In order to better 
evaluate coupling between electronic and nuclear structure and phonons, higher resolution INS data is required that can be rigorously compared with 
ab initio calculations.\\
In this context, we present a combined theoretical and experimental investigation of the lattice dynamics of BaFe$_2$As$_2$, including coupling to 
electronic and spin degrees of freedom. We report higher resolution INS spectra, above and below the phase transition temperature, which complement 
our published data. The generalised vibrational density of states is calculated for both phases using ab initio methods and allows the atomic contributions to 
the phonons to be evaluated. Phonon calculations have been repeated on experimentally determined crystal structures and on relaxed geometries, including 
spin polarisation, in order to identify critical, structural degrees of freedom. Finally the effect of phonons on the spin-resolved, electronic density 
of states is investigated.
\begin{table*}[htbp]
\centering
\caption{Labels for the different ab initio simulations presently reported.}
\begin{tabular}{ccc}
\hline
\hline
Label & Structural Data & Phase \\
\hline
S1 & experimental & tetragonal \\
S2 & experimental & orthorhombic \\
S31 & fully relaxed & orthorhombic \\
S32 & partially optimized (only z(As) is optimized) & orthorhombic \\
S41 & fully relaxed + magnetic ordering & orthorhombic \\
S42 & partially optimized (only z(As) is optimized) + magnetic oredering & orthorhombic \\
\hline
\hline
\end{tabular}
\end{table*}
\section{Experimental details}
The neutron spectroscopy measurements were performed on the cold time-of-flight spectrometer IN6 (Institut Laue Langevin, Grenoble, France), in 
up-scattering mode using an incident wavelength of 5.1 \AA\ and time-focussing at 4 meV giving very good resolution below 20 meV. Compared to our 
previous work~\cite{mittal}, where the thermal neutron time-of-flight spectrometer IN4 was used in down-scattering, we have better resolution but cannot 
access the full spectrum below a temperature of 140 K. The measurements from the two instruments are in this respect fully complementary and both 
structural phases can just be measured on IN6. Two grams of sample of polycrystalline  BaFe$_2$As$_2$ were prepared and characterized as reported 
elsewhere~\cite{rotter}. The sample was held in a thin Al-foil container and thermalized within a nitrogen cryo-loop for measurements at 140 and 300 K.
The container scattering was subtracted and the detectors calibrated with a vanadium standard. After correction for energy dependent detector efficiency, 
the one-phonon, generalized phonon DOS (GDOS) was obtained from the angle-integrated data (17 to 114$^\circ$) using the incoherent 
approximation~\cite{Schober-PRB-1997}. Multi-phonon terms to the scattering were determined self-consistently in an iterative procedure. 
Figure 1 (bottom panel) shows the measured spectra. 
The temperature of the structural transition is reported to be 140 K~\cite{rotter} or 142 K~\cite{huang}, with peak intensities starting to change significantly at temperatures as high as 150 K. 
The 140 K spectrum is therefore measured very close to the transition temperature but it shows some notable differences with the 300 K spectrum, like the hardening of the highest frequency vibrations and the band at 15 meV. These trends tend to be reproduced by the calculations (see below). Bigger differences between the spectra of the two structural phases are not expected since the data measured over a wide temperature range, at lower resolution, on IN4 did not reveal any marked changes~\cite{mittal}.
\section{Computational details}
Electronic and ionic first principle calculations were performed using the projector-augmented wave (PAW) formalism~\cite{paw} of the 
Kohn-Sham DFT~\cite{ks1,ks2}  at the generalized gradient approximation level (GGA), implemented in the Vienna ab initio simulation package 
(VASP)~\cite{vasp1,vasp2}. The GGA was formulated by the Perdew-Burke-Ernzerhof (PBE)~\cite{pbe1,pbe2} density functional. The Gaussian broadening 
technique was adopted and all results are well converged with respect to {\it k}-mesh and energy cutoff for the plane wave expansion. \\
Experimentally refined crystallographic data in both the low-temperature (orthorhombic) and high-temperature (tetragonal) phases were taken from 
reference~\cite{rotter}. These structures were used to calculate the GDOS and dispersion relations for both the tetragonal and orthorhombic phases. 
Hereafter, the corresponding simulations are labelled S1 and S2, respectively. The orthorhombic phase was then taken for a case study for which full 
(S31) and partial (S32) structural relaxation were performed. By partial relaxation we mean only the As z position 
is optimized. The effect of the observed magnetic ordering~\cite{huang} in the orthorhombic phase was taken into account in further full (S41) 
and partial (S42) optimizations. Table I provides a summary of these simulations. In the magnetic structure the Fe moments are parallel to the longer 
a-axis. Spins are aligned ferromagnetically along the shorter b-axis and antiferromagnetically along the a and c axes~\cite{huang}.\\
In the lattice dynamics calculations, in order to determine all inter-atomic force constants, the supercell approach has been adopted~\cite{direct}. 
Therefore, the single cells were used to construct (3*a, 3*b, c) and (2*a, 2*b, c) supercells for the tetragonal and orthorhombic cases, respectively, 
a and b being the shorter cell axes. The former contains 18 formula-units (90 atoms) and the latter contains 16 formula-units (80 atoms). Total energies 
and inter-atomic forces were calculated for the 12 and 18 structures (in the tetragonal and orthorhombic supercells, respectively) resulting from 
individual displacements of the three symmetry inequivalent atoms, along the three inequivalent Cartesian directions ($\pm$x, $\pm$y and $\pm$z). The 15 
phonons branches corresponding to the 5 atoms in the primitive cell, were extracted in subsequent calculations using the Phonon software~\cite{phonon}. 
\section{Results}
\begin{table*}[htbp]
\centering
\caption{Experimental and optimized lattice parameters and fractional As z-coordinates for both structural phases. NM, F and M$_{obs}$ label non-magnetic, ferromagnetic and observed 
magnetic ordering. Values between brackets indicate optimized As z position obtained from partial structural relaxations.}
\begin{tabular}{c|ccc|ccc}
\hline
\hline
       &     & Orthorhombic phase  &      &                  & Tetragonal phase &              \\
\hline
       & Exp~\cite{rotter}     & M$_{obs}$  & NM/F     & Exp~\cite{rotter} & M$_{obs}$ & NM/F    \\
a      & 5.6146  &    5.6611  &  5.6026  & 3.9625   & 3.9746    & 3.9553  \\
b      & 5.5742  &    5.5940  &  5.5998  & 3.9625   & 3.9746    & 3.9553  \\
c      & 12.9453 &   12.8793  &  12.6302 & 13.0168 & 12.7154   & 12.6849  \\
z(As)  & 0.3538  &    0.3512 (0.3510) & 0.3456 (0.3446) & 0.3545 & 0.3469 (0.3467) & 0.3455 (0.3442) \\
\hline
\hline
\end{tabular}
\end{table*}
\subsection{Generalized phonon density of states and dispersion relations}
Figure 1 shows the calculated dispersion curves and the GDOS for both the tetragonal (high-temperature: S1) and orthorhombic (low-temperature: S2) 
phases using the experimental structures~\cite{rotter}.
In order to compare with experimental data, the calculated GDOS was determined as the sum of the partial DOS weighted by the atomic scattering cross 
sections and masses: GDOS=$\sum_{i} \frac{\sigma_i}{M_i}$ pDOS$_i$ ($\frac{\sigma_i}{M_i}$ = 0.21 (Fe), 0.07 (As),  0.02 (Ba); i=\{Fe, As, Ba\}).
Both the full (GDOS) and partial gdos (pDOS) are included in figure 1. The
comparison of the calculated and measured spectra is generally good. The acoustic region extends in both phases to about 9 meV.
The lowest lying optic modes give rise to intensity starting at about 12 meV. 
As can be seen from the phonon relations, the corresponding dispersion sheets are not flat.
While at the $\Gamma$-point these low frequency, optic modes correspond to rigid displacements of the FeAs layers with respect to the Ba spacers,
distortions within the FeAs layers are present in the zone boundary modes. $\Gamma$-point modes above 18 meV distort the FeAs layers and the Raman active ones are illustrated in Figure 2(c).
The calculated vibrational spectrum is clearly separated into three bands via two pseudo-gaps, which arise from the strongly dispersive sheets 
in the regions around 20 and 28 meV.\\
The calculated phonon dispersion is sensitive to structural details, particularly for the optic bands in the intermediate frequency 
range from 10 - 20 meV. Flat bands lead to a clear two-peak structure in the GDOS at 13 and 18 meV in the orthorhombic phase. These peaks are found at 
13 and 16 meV in the tetragonal phase in good agreement with peaks in the corresponding experimental data. In particular, the experimentally observed 
peak at 16 meV in the orthorhombic phase softens to 15 meV in the tetragonal phase. Otherwise at the highest frequencies, bands at 24 and 29 meV in the 
tetragonal phase harden to 25 and 31 meV in the orthorhombic phase, following the trends observed experimentally. It is the new, higher resolution INS 
data that allows the validity of the phonon calculations, with respect to the structural phase transition, to be established.\\
However there is a peak observed at 20-21 meV which is absent in the calculated spectra for both phases. This constitutes the only significant 
discrepancy between experiment and calculation. As can be seen in the partial DOS, the dispersion branches in the region under question represent 
motions of Fe and As.
\subsection{Effect of structural relaxation}
A generally good description of the phonon spectra has been obtained by using the experimental structures for the ab initio calculations. However, 
experimental structures are not exactly the equilibrium structures that are typically used for phonon calculations. Phonon calculations have therefore 
been repeated in the orthorhombic phase for which full (S31 - atomic $z$ co-ordinate of As atom plus lattice constants) and partial 
(S32 - atomic co-ordinate only) structural relaxation were performed. In Figure 2(a) the corresponding spectra are compared with the calculation 
performed on the experimental cell (S2) and the experimental spectrum. The S31 and S32 spectra are almost identical meaning that the lattice constants 
alone have only a rather minor influence on the phonon spectra. The lattice constant which changes most on full relaxation is the c parameter, which 
determines the spacing between Ba and Fe-As layers (see Table II). The highest frequency modes, above 24 meV react most significantly to these structural 
changes. For example, the band at 25 meV splits in both S31 and S32, whereas a much weaker splitting is observed in the experimental data. The agreement 
with the experimental data is worse than for the calculation performed on the experimental structure, due to the hardening of the highest frequency 
vibrations. The only internal structural parameter is the z coordinate of the As atom, which deviates by ~0.1 $\AA$ from the experimental value upon relaxation. This is cause of the 
disagreement at high frequency.
\begin{figure*}
  \centerline{\hbox{ \hspace{-0.35in}
    \epsfxsize=3.6in
    \epsffile{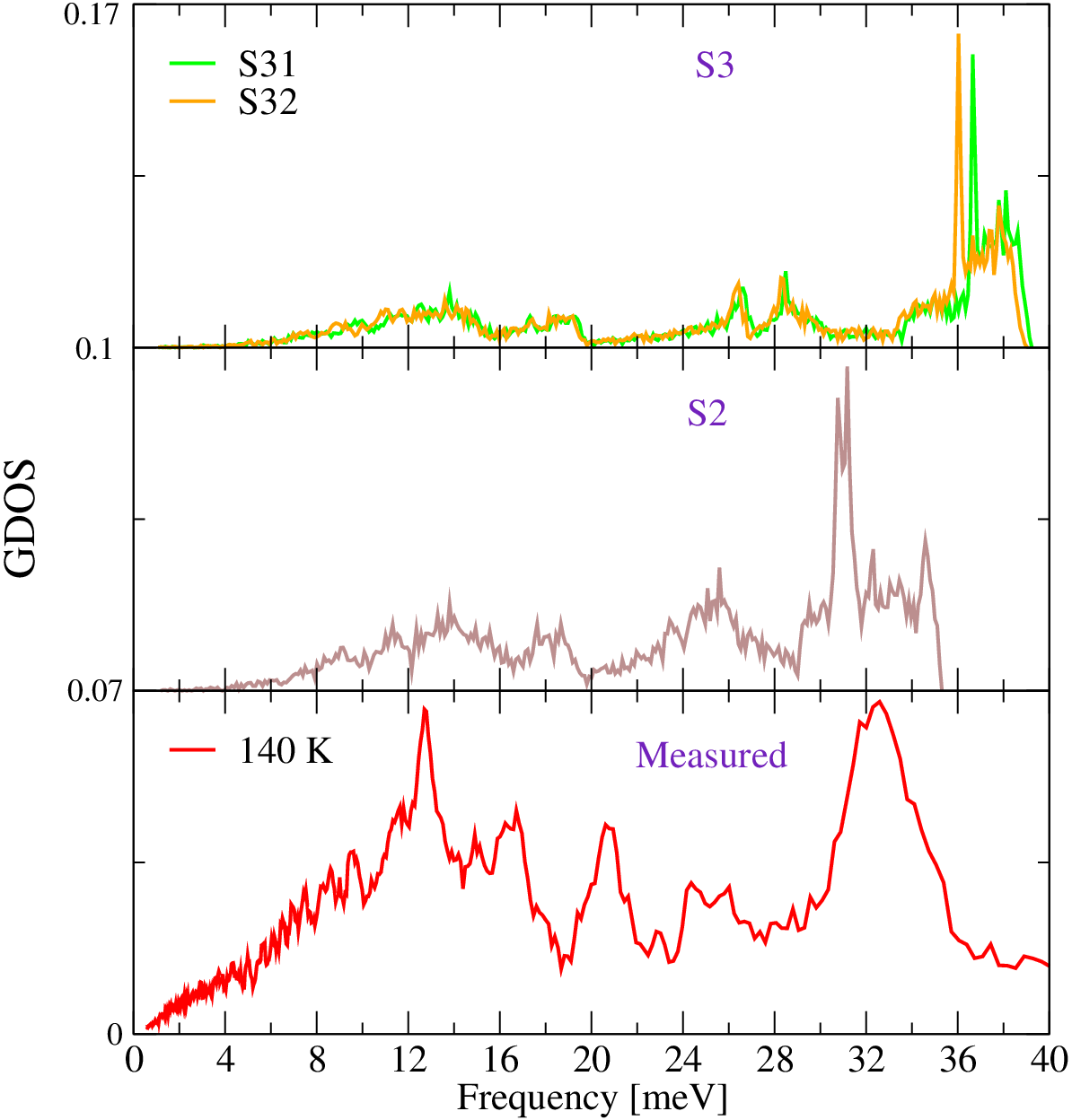}
    \hspace{0.25in}
    \epsfxsize=3.6in
    \epsffile{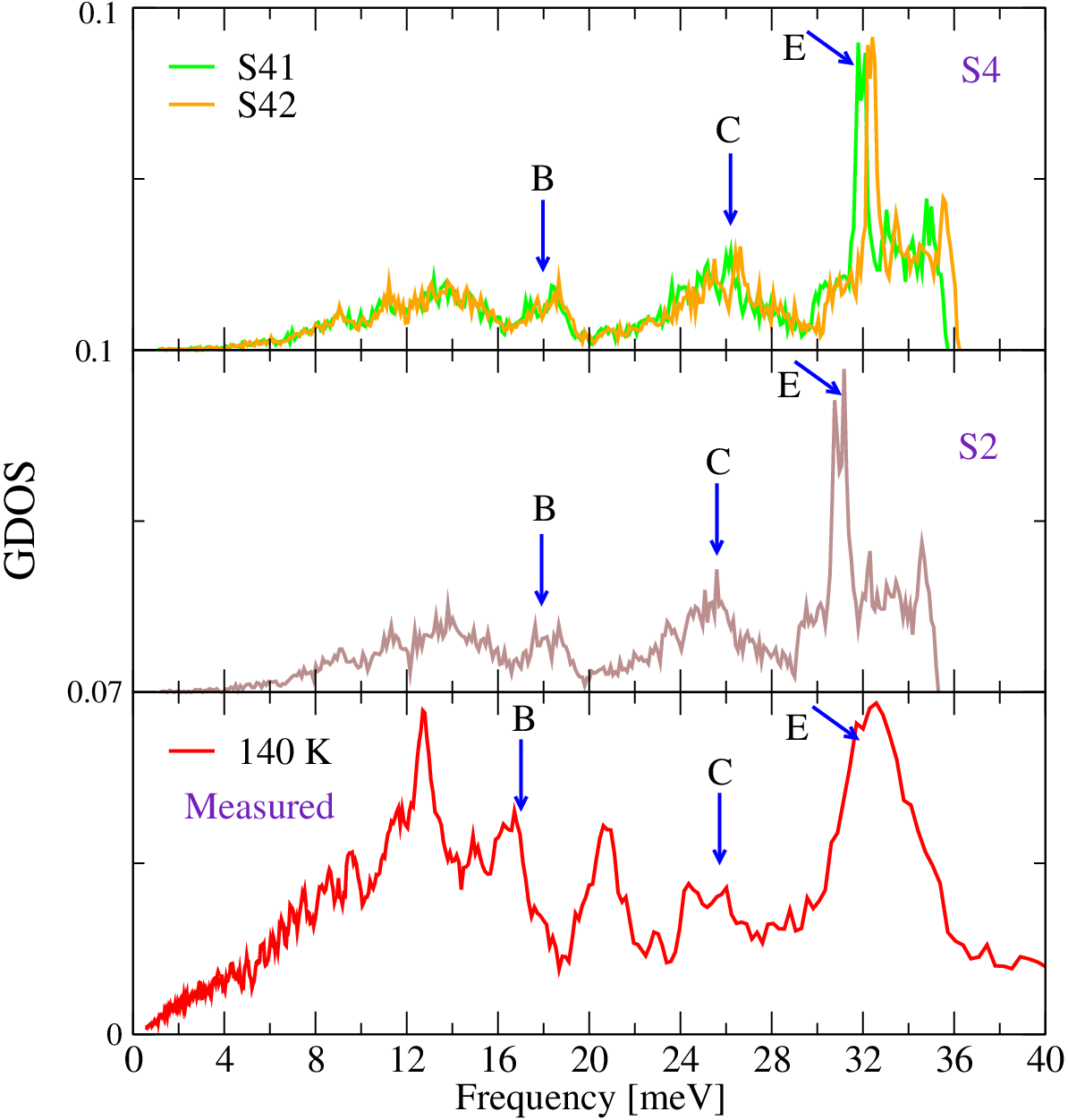}
    }
  }
  \hbox{\hspace{1.4in} (a) \hspace{4.0in} (b)}
 \centerline{\hbox{ \hspace{0.0in}
    \epsfxsize=3.6in
    \epsffile{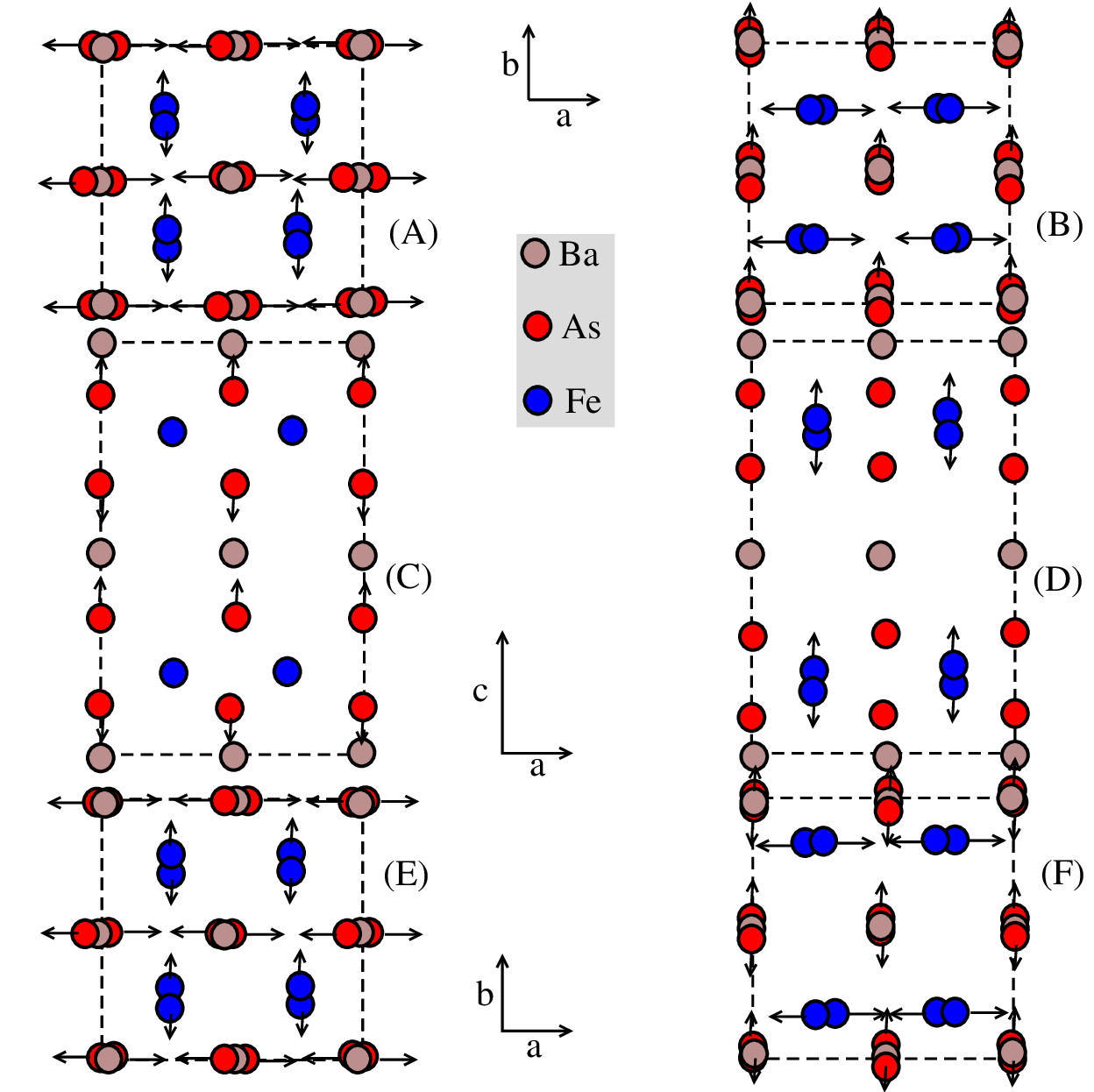}
    }
  }
  \hbox{\hspace{3.5in} (c)}
   \caption{(Color online) Measured and simulated GDOS in the orthorhombic phase (a) for relaxed non-magnetic structures (S31 and S32) and (b) for relaxed structures including the known magnetic ordering (S41 and S42). In both cases, the calculation based on the experimental, orthorhombic structure S2 is shown for comparison. Labels B, C and E in (b) indicate 
specific phonon modes which will be considered to discusse their effect on the electronic structure of BaFe$_2$As$_2$(cf Figure 3). The full displacement 
patterns of the six Raman active modes (A: 18 meV, B: 18 meV, C: 26 meV, D: 29 meV, E: 33 meV and F: 34 meV ), in the orthorhombic phase, are illustrated in (c).}
\end{figure*}
\subsection{Effect of magnetic ordering}
Further full (S41) and partial (S42) structural relaxation in the orthorhombic phase were performed taking into account the observed magnetic 
ordering~\cite{huang} (see Figure 2(b)). This leads to a description of the GDOS which is significantly better than that shown in Figure 2(a) for 
the optimised, non-magnetic structures and is comparable to that obtained with the experimental structure. This is appropriate from a theoretical 
point of view as the phonon calculations should start from a fully relaxed structure. The optimised c-axis is within 0.5\% of the experimental value 
and the As z-coordinate is close to the experimental value (see Table II). In addition, the calculated, orthorhombic distortion is strongest for 
the observed magnetic structure and closely matches the experimental value. Magnetic ordering of spin states in Fe cations is therefore important for 
calculating FeAs bonding and giving the best, calculated vibrational density of states. The data in Table II also reveals that, from a structural point 
of view, the ferromagnetic case is similar to the non-magnetic case. Also, including the correct magnetic ordering M$_{obs}$~\cite{huang} results in 
a volume expansion, which is of interest in light of the apparent pressure-induced superconductivity in the parent compounds of the FeAs-based 
superconductors~\cite{colaps, pressure}.
\begin{figure*}[htbp]
\begin{center}
\epsfig{file=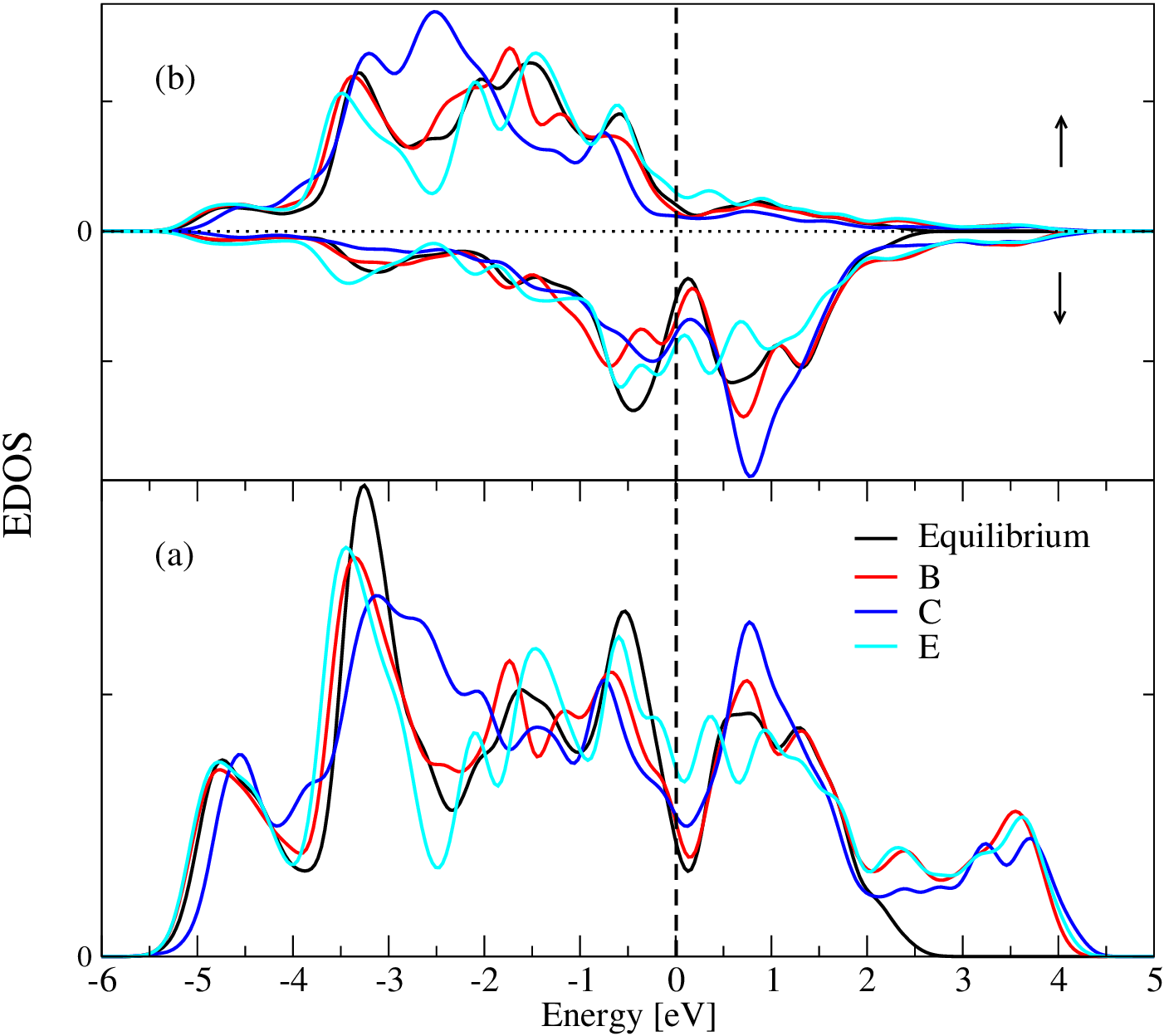, width=12 cm}
\end{center}
\caption{(Color online) Bottom panel (a): total eDOS of the unperturbed (Equilibrium), and the structurally perturbed orthorhombic phase of BaFe$_2$As$_2$. Total eDOS 
of spin-down channel is identical to the up-channel. Top panel (b): the corresponding spin-resolved eDOS of the d-electrons per Fe cation. Labels B, C 
and E indicate phonon-perturbed equilibrium structures (cf Figure 2(b) and Figure 2(c); B: 18 meV, C: 26 meV and E: 32 meV).}
\end{figure*}
\subsection{Effect of phonons on the electron density of states}
To gain insight into the effect of phonons on electronic structure,
we have calculated the electronic DOS (eDOS) for frozen phonon patterns using the recently observed magnetic structure in the orthorhombic 
phase~\cite{huang}. The mean amplitude was chosen as $\sim$ 0.05 \AA , which corresponds approximately to a single occupation of the mode. 
The results are shown in Figure 3(a) for the three Gamma point modes that most strongly perturb the equilibrium electronic structure. These modes are 
Raman active and are labelled B, C and E in Figure 2(c) and have frequencies 18, 26 and 32 meV. The main features of the eDOS for the unperturbed 
structure are in good agreement with other electronic structure calculations of the tetragonal phase~\cite{edos1,edos2,edos3}. 
The bands close to the Fermi level and in particular the eDOS at the Fermi level depend strongly on the phonon distortion via the d-bands of Fe. As the 
Fermi level of the equilibrium structure is found at the minimum of a rather steep-walled valley of the eDOS, phonon modes that displace the Fermi level 
only slightly, create a large fluctuation of the electron density at the Fermi-level.\\
In view of the importance of magnetic ordering on the equilibrium structure of BaFe$_2$As$_2$, we have also investigated the spin-resolved eDOS of the 
d-electrons of the Fe cation (see Figure 3(b)). Complex changes in the eDOS are captured in the magnetic moment of the Fe cation. The calculated magnetic 
moment for the equilibiurm structure is 1.94 $\mu_B$. 
This value is much larger than the observed magnetic moment (~1 $\mu_B$) as reported by other authors~\cite{lebegue} and explained in terms of frustrating 
magnetic interactions reducing the observed moment (see ~\cite{huang} and references therein). On the computational side, standard settings, notably for 
band occupation at the Fermi level, tend to result in more localised and therefore bigger moments. Settings could be tuned to give a reduced magnetic 
moment but here we simply want to use the equilibrium magnetic moment as a reference value to quantify the effect of structural distortions. 
Mode B increases the moment slightly to 2.07 $\mu_B$, mode C increases it to 2.55 $\mu_B$ and mode E decreases it to 1.58 $\mu_B$. From 
Figure 3(b) it is seen that modes B and C cause a decrease in the spin-down population and an increase in the spin-up population and therefore an 
increase in the magnetic moment. On the other hand, mode E causes a rocking of the spin population from the majority, up-channel to the minority 
down-channel, decreasing the magnetic moment. Mode E tends to reduce the magnetic ordering. For all 3 modes, the Fermi level changes by less 0.5\%.
\section{Conclusion and Perspectives}
Small variations in the GDOS between the tetragonal and orthorhombic phases are observed in the new, high resolution, experimental data and these are 
generally reproduced by the calculations. The variations in the distances between Fe atoms and between Fe and As atoms are less than 
0.01 \AA~\cite{rotter}, revealing the sensitivity of phonons to subtle structural changes. Phonon calculations performed on the experimental structure 
and on the fully relaxed structure, including the observed magnetic ordering, best match the experimental spectra. Non-magnetic and ferromagnetic, 
relaxed structures show bigger discrepancies with the measured spectra. The key structural parameter is the z-coordinate of the As atom which governs 
the distortion of the Fe-As layer. Phonon modes that modulate the Fe-As layer cause the most significant changes in the eDOS at the Fermi level, which 
indicate the extent of electron-phonon coupling. In addition the magnetic moment and spin-resolved eDOS of the Fe cation vary significantly with the 
phonon displacements, demonstrating spin-phonon coupling, which is relevant to the search for a composite mechanism for high temperature 
superconductivity that goes beyond simple, electron-phonon coupling. \\
The main short-coming of the calculation is the inability to reproduce the observed, experimental peak at 20 meV. The calculation predicts in this 
region highly dispersed branches leading to a pseudo-gap in the density-of-states. Experimentally, the $Q$-dependence of this peak is similar to that 
of other phonon bands. There is, therefore, no experimental indication that we are dealing with a magnetic excitation, especially at the temperatures of 
these measurements. In recent work by Reznik and coworkers~\cite{reznik} combining ab initio, linear response, phonon 
calculations and inelastic X-ray scattering, 
the peak close at 20 meV is observed experimentally but not reproduced computationally.\\
We have also tried to generate spectral intensity at 20 meV within the framework of our phonon calculations. This is achieved by trying to fit 
inter-atomic force constants to change the frequencies of selected phonons at particular wave-vectors. The only phonon that could be significantly 
modified in this way was a mode at 24 meV at the zone boundary (0,0,0.5), but the change in spectral intensity is small. This effect is however consistent with the sensitivity that we have demonstrated of the 
phonons to the z-coordinate of the As atom. The importance of the c-axis phonons in FeAs based superconductors has also been highlighted recently~\cite{reznik}. \\
However, if we believe that modern DFT codes correctly reproduce the main spectral features in, structurally, rather simple systems like 
BaFe$_2$As$_2$, then we have no clear explanation for this peak within the phonon model implemented here. Therefore, any statement on the 
origin of the 20  meV peak without further experiments has to remain speculative. Strong, non-conventional electron-phonon coupling has been proposed 
in the literature~\cite{Eschrig} for LaOFeAs. The possibility of a singularity close to 20 meV is evoked on the basis of gap openings at the Fermi 
surface. The same physical arguments should hold for BaFe$_2$As$_2$ as only the FeAs-layers are concerned.\\
Finally we plan to extend our measurements and calculations to other systems in this family of compounds, to determine whether the anomaly at 20 
meV is a general feature or whether it is specific to the Ba system. In this context, we are motivated by initial work on the strontium system for 
which we obtain excellent agreement between measured and calculated Raman frequencies.


\begin{thebibliography}{31}
\expandafter\ifx\csname natexlab\endcsname\relax\def\natexlab#1{#1}\fi
\expandafter\ifx\csname bibnamefont\endcsname\relax
  \def\bibnamefont#1{#1}\fi
\expandafter\ifx\csname bibfnamefont\endcsname\relax
  \def\bibfnamefont#1{#1}\fi
\expandafter\ifx\csname citenamefont\endcsname\relax
  \def\citenamefont#1{#1}\fi
\expandafter\ifx\csname url\endcsname\relax
  \def\url#1{\texttt{#1}}\fi
\expandafter\ifx\csname urlprefix\endcsname\relax\def\urlprefix{URL }\fi
\providecommand{\bibinfo}[2]{#2}
\providecommand{\eprint}[2][]{\url{#2}}

\bibitem[{\citenamefont{Kamihara et~al.}(2006)\citenamefont{Kamihara,
  Hiramatsu, Hirano, Kawamura, Yanagi, Kamiya, and Hosono}}]{super1}
\bibinfo{author}{\bibfnamefont{Y.}~\bibnamefont{Kamihara}},
  \bibinfo{author}{\bibfnamefont{H.}~\bibnamefont{Hiramatsu}},
  \bibinfo{author}{\bibfnamefont{M.}~\bibnamefont{Hirano}},
  \bibinfo{author}{\bibfnamefont{R.}~\bibnamefont{Kawamura}},
  \bibinfo{author}{\bibfnamefont{H.}~\bibnamefont{Yanagi}},
  \bibinfo{author}{\bibfnamefont{T.}~\bibnamefont{Kamiya}}, \bibnamefont{and}
  \bibinfo{author}{\bibfnamefont{H.}~\bibnamefont{Hosono}},
  \bibinfo{journal}{J.~Am.~Chem.~Soc.} \textbf{\bibinfo{volume}{128}},
  \bibinfo{pages}{10012} (\bibinfo{year}{2006}).

\bibitem[{\citenamefont{Liang et~al.}(2007)\citenamefont{Liang, Che, Yang,
  Tian, Xiao, Lu, Li, and Li}}]{super2}
\bibinfo{author}{\bibfnamefont{C.~Y.} \bibnamefont{Liang}},
  \bibinfo{author}{\bibfnamefont{R.~C.} \bibnamefont{Che}},
  \bibinfo{author}{\bibfnamefont{H.~X.} \bibnamefont{Yang}},
  \bibinfo{author}{\bibfnamefont{H.~F.} \bibnamefont{Tian}},
  \bibinfo{author}{\bibfnamefont{R.~J.} \bibnamefont{Xiao}},
  \bibinfo{author}{\bibfnamefont{J.~B.} \bibnamefont{Lu}},
  \bibinfo{author}{\bibfnamefont{R.}~\bibnamefont{Li}}, \bibnamefont{and}
  \bibinfo{author}{\bibfnamefont{J.~Q.} \bibnamefont{Li}},
  \bibinfo{journal}{Supercond.~Sci.~Technol.} \textbf{\bibinfo{volume}{20}},
  \bibinfo{pages}{687} (\bibinfo{year}{2007}).

\bibitem[{\citenamefont{Watanabe et~al.}(2007)\citenamefont{Watanabe, Yanagi,
  Kamiya, Kamihara, Hiramatsu, Hirano, and Hosono}}]{super3}
\bibinfo{author}{\bibfnamefont{T.}~\bibnamefont{Watanabe}},
  \bibinfo{author}{\bibfnamefont{H.}~\bibnamefont{Yanagi}},
  \bibinfo{author}{\bibfnamefont{T.}~\bibnamefont{Kamiya}},
  \bibinfo{author}{\bibfnamefont{Y.}~\bibnamefont{Kamihara}},
  \bibinfo{author}{\bibfnamefont{H.}~\bibnamefont{Hiramatsu}},
  \bibinfo{author}{\bibfnamefont{M.}~\bibnamefont{Hirano}}, \bibnamefont{and}
  \bibinfo{author}{\bibfnamefont{H.}~\bibnamefont{Hosono}},
  \bibinfo{journal}{Inorg.~Chem} \textbf{\bibinfo{volume}{46}},
  \bibinfo{pages}{7719} (\bibinfo{year}{2007}).

\bibitem[{\citenamefont{Kamihara et~al.}(2008)\citenamefont{Kamihara, Watanabe,
  Hirano, and Hosono}}]{super4}
\bibinfo{author}{\bibfnamefont{O.}~\bibnamefont{Kamihara}},
  \bibinfo{author}{\bibfnamefont{T.}~\bibnamefont{Watanabe}},
  \bibinfo{author}{\bibfnamefont{M.}~\bibnamefont{Hirano}}, \bibnamefont{and}
  \bibinfo{author}{\bibfnamefont{H.}~\bibnamefont{Hosono}},
  \bibinfo{journal}{J.~Am.~Chem.~Soc.} \textbf{\bibinfo{volume}{130}},
  \bibinfo{pages}{3296} (\bibinfo{year}{2008}).

\bibitem[{\citenamefont{Hunte et~al.}(2008)\citenamefont{Hunte, Jaroszynski,
  Gurevich, Larbalestier, Jin, Sefat, McGuire, Sales, Christen, and
  Mandrus}}]{super5}
\bibinfo{author}{\bibfnamefont{F.}~\bibnamefont{Hunte}},
  \bibinfo{author}{\bibfnamefont{J.}~\bibnamefont{Jaroszynski}},
  \bibinfo{author}{\bibfnamefont{A.}~\bibnamefont{Gurevich}},
  \bibinfo{author}{\bibfnamefont{D.~C.} \bibnamefont{Larbalestier}},
  \bibinfo{author}{\bibfnamefont{R.}~\bibnamefont{Jin}},
  \bibinfo{author}{\bibfnamefont{A.~S.} \bibnamefont{Sefat}},
  \bibinfo{author}{\bibfnamefont{M.~A.} \bibnamefont{McGuire}},
  \bibinfo{author}{\bibfnamefont{B.~C.} \bibnamefont{Sales}},
  \bibinfo{author}{\bibfnamefont{D.~K.} \bibnamefont{Christen}},
  \bibnamefont{and} \bibinfo{author}{\bibfnamefont{D.}~\bibnamefont{Mandrus}},
  \bibinfo{journal}{Nature.} \textbf{\bibinfo{volume}{453}},
  \bibinfo{pages}{903} (\bibinfo{year}{2008}).

\bibitem[{\citenamefont{Alireza et~al.}()\citenamefont{Alireza, Gillett, Ko,
  Sebastian, and Lonzarich}}]{Alireza-2008}
\bibinfo{author}{\bibfnamefont{P.~L.} \bibnamefont{Alireza}},
  \bibinfo{author}{\bibfnamefont{J.}~\bibnamefont{Gillett}},
  \bibinfo{author}{\bibfnamefont{Y.~T.~C.} \bibnamefont{Ko}},
  \bibinfo{author}{\bibfnamefont{S.~E.} \bibnamefont{Sebastian}},
  \bibnamefont{and} \bibinfo{author}{\bibfnamefont{G.~G.}
  \bibnamefont{Lonzarich}}, \emph{\bibinfo{title}{arxiv.org:0807.1896 (2008).}}

\bibitem[{\citenamefont{Singh and Du}(2008)}]{phon1}
\bibinfo{author}{\bibfnamefont{D.~J.} \bibnamefont{Singh}} \bibnamefont{and}
  \bibinfo{author}{\bibfnamefont{M.-H.} \bibnamefont{Du}},
  \bibinfo{journal}{Phys.~Rev.~Lett.} \textbf{\bibinfo{volume}{100}},
  \bibinfo{pages}{237003} (\bibinfo{year}{2008}).

\bibitem[{\citenamefont{Boeri et~al.}(2008)\citenamefont{Boeri, Dolgov, and
  Golubov}}]{phon2}
\bibinfo{author}{\bibfnamefont{L.}~\bibnamefont{Boeri}},
  \bibinfo{author}{\bibfnamefont{O.~V.} \bibnamefont{Dolgov}},
  \bibnamefont{and} \bibinfo{author}{\bibfnamefont{A.~A.}
  \bibnamefont{Golubov}}, \bibinfo{journal}{Phys.~Rev.~Lett.}
  \textbf{\bibinfo{volume}{101}}, \bibinfo{pages}{026403}
  (\bibinfo{year}{2008}).

\bibitem[{\citenamefont{Eschrig}()}]{Eschrig}
\bibinfo{author}{\bibfnamefont{H.}~\bibnamefont{Eschrig}},
  \emph{\bibinfo{title}{arxiv.org:0804.0186 (2008).}}

\bibitem[{\citenamefont{Pickett}(2002)}]{Pickett-Nature-2002}
\bibinfo{author}{\bibfnamefont{W.}~\bibnamefont{Pickett}},
  \bibinfo{journal}{Nature.} \textbf{\bibinfo{volume}{418}},
  \bibinfo{pages}{733} (\bibinfo{year}{2002}).

\bibitem[{\citenamefont{Boothroyd}(2008)}]{Boothroyd-PRB}
  \bibinfo{author}{\bibfnamefont{R.~A.}~\bibnamefont{Ewings}},
  \bibinfo{author}{\bibfnamefont{T.~G.}~\bibnamefont{Perring}},
  \bibinfo{author}{\bibfnamefont{R.~I.}~\bibnamefont{Bewley}},
  \bibinfo{author}{\bibfnamefont{T.}~\bibnamefont{Guidi}},
  \bibinfo{author}{\bibfnamefont{M.~J.}~\bibnamefont{Pitcher}},
  \bibinfo{author}{\bibfnamefont{D.~R.}~\bibnamefont{Parker}},
  \bibinfo{author}{\bibfnamefont{S.~J.}~\bibnamefont{Clarke}}, \bibnamefont{and}
  \bibinfo{author}{\bibfnamefont{A.~T. }~\bibnamefont{Boothroyd}},
  \emph{\bibinfo{title}{arxiv.org:0808.2836 (2008).}}

\bibitem[{\citenamefont{Renker et~al.}(2002)\citenamefont{Christianson}}]{christianson}
  \bibinfo{author}{\bibfnamefont{A. D.}~\bibnamefont{Christianson}},
  \bibinfo{author}{\bibfnamefont{M. D.}~\bibnamefont{Lumsden}},
  \bibinfo{author}{\bibfnamefont{O.}~\bibnamefont{Delaire}},
  \bibinfo{author}{\bibfnamefont{M. B.}~\bibnamefont{Stone}},
  \bibinfo{author}{\bibfnamefont{D. L.}~\bibnamefont{Abernathy}},
  \bibinfo{author}{\bibfnamefont{M. A.}~\bibnamefont{McGuire}},
  \bibinfo{author}{\bibfnamefont{A. S.}~\bibnamefont{Sefat}},
  \bibinfo{author}{\bibfnamefont{R.}~\bibnamefont{Jin}},
  \bibinfo{author}{\bibfnamefont{B. C.}~\bibnamefont{Sales}},
  \bibinfo{author}{\bibfnamefont{D.}~\bibnamefont{Mandrus}},
  \bibinfo{author}{\bibfnamefont{E. D.}~\bibnamefont{Mun}},
  \bibinfo{author}{\bibfnamefont{P. C.}~\bibnamefont{Canfield}},
  \bibinfo{author}{\bibfnamefont{J. Y. Y.}~\bibnamefont{Lin}},
  \bibinfo{author}{\bibfnamefont{M.}~\bibnamefont{Lucas}},
  \bibinfo{author}{\bibfnamefont{M.}~\bibnamefont{M. Kresch}},
  \bibinfo{author}{\bibfnamefont{J. B.}~\bibnamefont{Keith}},
  \bibinfo{author}{\bibfnamefont{B.}~\bibnamefont{Fultz}},
  \bibinfo{author}{\bibfnamefont{E. A.}~\bibnamefont{Goremychkin}}, \bibnamefont{and}
  \bibinfo{author}{\bibfnamefont{R. J.}~\bibnamefont{McQueeney}},
  \bibinfo{journal}{Phys.~Rev.~Lett.} \textbf{\bibinfo{volume}{101}},
  \bibinfo{pages}{157004} (\bibinfo{year}{2008}).

\bibitem[{\citenamefont{Mittal et~al.}(2008)\citenamefont{Mittal, Su, Rols, Chatterji, Chaplot, Schober, Rotter, Johrendt, and Brueckel}}]{mittal}
\bibinfo{author}{\bibfnamefont{R.}~\bibnamefont{Mittal}},
  \bibinfo{author}{\bibfnamefont{Y.}~\bibnamefont{Su}},
  \bibinfo{author}{\bibfnamefont{S.}~\bibnamefont{Rols}},
  \bibinfo{author}{\bibfnamefont{T.}~\bibnamefont{Chatterji}},
  \bibinfo{author}{\bibfnamefont{S.~L.} \bibnamefont{Chaplot}},
  \bibinfo{author}{\bibfnamefont{H.}~\bibnamefont{Schober}},
  \bibinfo{author}{\bibfnamefont{M.}~\bibnamefont{Rotter}},
  \bibinfo{author}{\bibfnamefont{D.}~\bibnamefont{Johrendt}}, \bibnamefont{and}
  \bibinfo{author}{\bibfnamefont{T.}~\bibnamefont{Brueckel}},
  \bibinfo{journal}{Phys.~Rev. {\em B}} \textbf{\bibinfo{volume}{78}},
  \bibinfo{pages}{104514} (\bibinfo{year}{2008}).

\bibitem[{\citenamefont{Singh}(2002)\citenamefont{Christianson}}]{singh}
  \bibinfo{author}{\bibfnamefont{D. J.}~\bibnamefont{Singh}}, \bibnamefont{and}
  \bibinfo{author}{\bibfnamefont{M. H.}~\bibnamefont{Du}},
  \bibinfo{journal}{Phys.~Rev.~Lett.} \textbf{\bibinfo{volume}{100}},
  \bibinfo{pages}{237003} (\bibinfo{year}{2008}).

\bibitem[{\citenamefont{Rotter et~al.}(2008)\citenamefont{Rotter, Tegel,
  Johrendt, Schellenberg, Hermes, and P\"ottgen}}]{rotter}
\bibinfo{author}{\bibfnamefont{M.}~\bibnamefont{Rotter}},
  \bibinfo{author}{\bibfnamefont{M.}~\bibnamefont{Tegel}},
  \bibinfo{author}{\bibfnamefont{D.}~\bibnamefont{Johrendt}},
  \bibinfo{author}{\bibfnamefont{I.}~\bibnamefont{Schellenberg}},
  \bibinfo{author}{\bibfnamefont{W.}~\bibnamefont{Hermes}}, \bibnamefont{and}
  \bibinfo{author}{\bibfnamefont{R.}~\bibnamefont{P\"ottgen}},
  \bibinfo{journal}{Phys.~Rev. {\em B}} \textbf{\bibinfo{volume}{78}},
  \bibinfo{pages}{020503} (\bibinfo{year}{2008}).

\bibitem[{\citenamefont{Schober et~al.}(1997)\citenamefont{Schober, T\"olle, Renker, Heid, and Gompf}}]{Schober-PRB-1997}
\bibinfo{author}{\bibfnamefont{H.}~\bibnamefont{Schober}},
  \bibinfo{author}{\bibfnamefont{A.}~\bibnamefont{T\"olle}},
  \bibinfo{author}{\bibfnamefont{B.}~\bibnamefont{Renker}},
  \bibinfo{author}{\bibfnamefont{R.}~\bibnamefont{Heid}}, \bibnamefont{and}
  \bibinfo{author}{\bibfnamefont{F.}~\bibnamefont{Gompf}},
  \bibinfo{journal}{Phys.~Rev. {\em B}} \textbf{\bibinfo{volume}{56}},
  \bibinfo{pages}{5937} (\bibinfo{year}{1997}).

\bibitem[{\citenamefont{Renker et~al.}(2002)\citenamefont{Christianson}}]{huang}
  \bibinfo{author}{\bibfnamefont{Q.}~\bibnamefont{Huang}},
  \bibinfo{author}{\bibfnamefont{Y.}~\bibnamefont{Qiu}},
  \bibinfo{author}{\bibfnamefont{W.}~\bibnamefont{Bao}},
  \bibinfo{author}{\bibfnamefont{M. A.}~\bibnamefont{Green}},
  \bibinfo{author}{\bibfnamefont{J. W.}~\bibnamefont{Lynn}},
  \bibinfo{author}{\bibfnamefont{Y. C.}~\bibnamefont{Gasparovic}},
  \bibinfo{author}{\bibfnamefont{T.}~\bibnamefont{Wu}},
  \bibinfo{author}{\bibfnamefont{G.}~\bibnamefont{Wu}}, \bibnamefont{and}
  \bibinfo{author}{\bibfnamefont{X. H.}~\bibnamefont{Chen}},
  \bibinfo{journal}{Phys.~Rev.~Lett.} \textbf{\bibinfo{volume}{101}},
  \bibinfo{pages}{257003} (\bibinfo{year}{2008}).

\bibitem[{\citenamefont{Bl\"ochl}(1994)}]{paw}
\bibinfo{author}{\bibfnamefont{P.~E.} \bibnamefont{Bl\"ochl}},
  \bibinfo{journal}{Phys.~Rev. {\em B}} \textbf{\bibinfo{volume}{50}},
  \bibinfo{pages}{17953} (\bibinfo{year}{1994}).

\bibitem[{\citenamefont{Hohenberg and Kohn}(1964)}]{ks1}
\bibinfo{author}{\bibfnamefont{P.}~\bibnamefont{Hohenberg}} \bibnamefont{and}
  \bibinfo{author}{\bibfnamefont{W.}~\bibnamefont{Kohn}},
  \bibinfo{journal}{Phys.~Rev.} \textbf{\bibinfo{volume}{136}},
  \bibinfo{pages}{864} (\bibinfo{year}{1964}).

\bibitem[{\citenamefont{Kohn and Sham}(1965)}]{ks2}
\bibinfo{author}{\bibfnamefont{W.}~\bibnamefont{Kohn}} \bibnamefont{and}
  \bibinfo{author}{\bibfnamefont{L.~J.} \bibnamefont{Sham}},
  \bibinfo{journal}{Phys.~Rev.} \textbf{\bibinfo{volume}{140}},
  \bibinfo{pages}{1133} (\bibinfo{year}{1965}).

\bibitem[{\citenamefont{Kresse and Furthm\"uller}(1996)}]{vasp1}
\bibinfo{author}{\bibfnamefont{G.}~\bibnamefont{Kresse}} \bibnamefont{and}
  \bibinfo{author}{\bibfnamefont{J.}~\bibnamefont{Furthm\"uller}},
  \bibinfo{journal}{Comput.~Mater.~Sci.} \textbf{\bibinfo{volume}{6}},
  \bibinfo{pages}{15} (\bibinfo{year}{1996}).

\bibitem[{\citenamefont{Kresse and Joubert}(1999)}]{vasp2}
\bibinfo{author}{\bibfnamefont{G.}~\bibnamefont{Kresse}} \bibnamefont{and}
  \bibinfo{author}{\bibfnamefont{D.}~\bibnamefont{Joubert}},
  \bibinfo{journal}{Phys.~Rev. {\em B}} \textbf{\bibinfo{volume}{59}},
  \bibinfo{pages}{1758} (\bibinfo{year}{1999}).

\bibitem[{\citenamefont{Perdew et~al.}(1996)\citenamefont{Perdew, Burke, and
  Ernzerhof}}]{pbe1}
\bibinfo{author}{\bibfnamefont{J.~P.} \bibnamefont{Perdew}},
  \bibinfo{author}{\bibfnamefont{K.}~\bibnamefont{Burke}}, \bibnamefont{and}
  \bibinfo{author}{\bibfnamefont{M.}~\bibnamefont{Ernzerhof}},
  \bibinfo{journal}{Phys.~Rev.~Lett.} \textbf{\bibinfo{volume}{77}},
  \bibinfo{pages}{3865} (\bibinfo{year}{1996}).

\bibitem[{\citenamefont{Perdew et~al.}(1997)\citenamefont{Perdew, Burke, and
  Ernzerhof}}]{pbe2}
\bibinfo{author}{\bibfnamefont{J.~P.} \bibnamefont{Perdew}},
  \bibinfo{author}{\bibfnamefont{K.}~\bibnamefont{Burke}}, \bibnamefont{and}
  \bibinfo{author}{\bibfnamefont{M.}~\bibnamefont{Ernzerhof}},
  \bibinfo{journal}{Phys.~Rev.~Lett.} \textbf{\bibinfo{volume}{78}},
  \bibinfo{pages}{1396} (\bibinfo{year}{1997}).

\bibitem[{\citenamefont{Parlinski et~al.}(1997)\citenamefont{Parlinski, Li, and Kawazoe}}]{direct}
\bibinfo{author}{\bibfnamefont{K.}~\bibnamefont{Parlinski}},
  \bibinfo{author}{\bibfnamefont{Z.-Q.} \bibnamefont{Li}}, \bibnamefont{and}
  \bibinfo{author}{\bibfnamefont{Y.}~\bibnamefont{Kawazoe}},
  \bibinfo{journal}{Phys.~Rev.~Lett.} \textbf{\bibinfo{volume}{78}},
  \bibinfo{pages}{4063} (\bibinfo{year}{1997}).

\bibitem[{\citenamefont{Parlinksi}()}]{phonon}
\bibinfo{author}{\bibfnamefont{K.}~\bibnamefont{Parlinksi}},
  \emph{\bibinfo{title}{Software phonon, 2003}}.

\bibitem[{\citenamefont{Kreyssig et~al.}()\citenamefont{Kreyssig, Green, Lee, Samolyuk, Zajdel, Lynn, Bud'ko, Torikachvili, Ni, Nandi et~al.}}]{colaps}
\bibinfo{author}{\bibfnamefont{A.}~\bibnamefont{Kreyssig}},
  \bibinfo{author}{\bibfnamefont{M.~A.} \bibnamefont{Green}},
  \bibinfo{author}{\bibfnamefont{Y.}~\bibnamefont{Lee}},
  \bibinfo{author}{\bibfnamefont{G.~D.} \bibnamefont{Samolyuk}},
  \bibinfo{author}{\bibfnamefont{P.}~\bibnamefont{Zajdel}},
  \bibinfo{author}{\bibfnamefont{J.~W.} \bibnamefont{Lynn}},
  \bibinfo{author}{\bibfnamefont{S.~L.} \bibnamefont{Bud'ko}},
  \bibinfo{author}{\bibfnamefont{M.~S.} \bibnamefont{Torikachvili}},
  \bibinfo{author}{\bibfnamefont{N.}~\bibnamefont{Ni}}, \bibnamefont{and}
  \bibinfo{author}{\bibfnamefont{S.}~\bibnamefont{Nandi}},
  \emph{\bibinfo{title}{arxiv.org:0807.3032 (2008).}}


\bibitem[{\citenamefont{Kreyssig et~al.}()\citenamefont{Kreyssig, Green, Lee, Samolyuk, Zajdel, Lynn, Bud'ko, Torikachvili, Ni, Nandi et~al.}}]{pressure}
\bibinfo{author}{\bibfnamefont{W.}~\bibnamefont{Yu}},
  \bibinfo{author}{\bibfnamefont{A.~A.} \bibnamefont{Aczel}},
  \bibinfo{author}{\bibfnamefont{T. J.}~\bibnamefont{Williams}},
  \bibinfo{author}{\bibfnamefont{S.~L.} \bibnamefont{Bud'ko}},
  \bibinfo{author}{\bibfnamefont{N.}~\bibnamefont{Ni}},
  \bibinfo{author}{\bibfnamefont{P.~C.} \bibnamefont{Canfield}}, \bibnamefont{and}
  \bibinfo{author}{\bibfnamefont{G. M.}~\bibnamefont{Luke}},
  \emph{\bibinfo{title}{arxiv.org:0811.2554 (2008).}}

\bibitem[{\citenamefont{Shein and Ivanovskii}()}]{edos1}
\bibinfo{author}{\bibfnamefont{I.~R.} \bibnamefont{Shein}} \bibnamefont{and}
  \bibinfo{author}{\bibfnamefont{A.~L.} \bibnamefont{Ivanovskii}},
  \emph{\bibinfo{title}{arxiv.org:0806.0750 (2008).}}

\bibitem[{\citenamefont{Ma et~al.}()\citenamefont{Ma, LU, and Xiang}}]{edos2}
\bibinfo{author}{\bibfnamefont{F.}~\bibnamefont{Ma}},
  \bibinfo{author}{\bibfnamefont{Z.-Y.} \bibnamefont{LU}}, \bibnamefont{and}
  \bibinfo{author}{\bibfnamefont{T.}~\bibnamefont{Xiang}},
  \emph{\bibinfo{title}{arxiv.org:0806.3526 (2008).}}

\bibitem[{\citenamefont{Singh}(2008)}]{edos3}
\bibinfo{author}{\bibfnamefont{D.~J.} \bibnamefont{Singh}},
  \bibinfo{journal}{Phys.~Rev. {\em B}} \textbf{\bibinfo{volume}{78}},
  \bibinfo{pages}{094511} (\bibinfo{year}{2008}).

\bibitem[{\citenamefont{lebegue}()\citenamefont{Ma, LU, and Xiang}}]{lebegue}
  \bibinfo{author}{\bibfnamefont{S.}~\bibnamefont{Leb\`egue}},
  \bibinfo{author}{\bibfnamefont{Z. P.}~\bibnamefont{Yin}}, \bibnamefont{and}
  \bibinfo{author}{\bibfnamefont{W. E.}~\bibnamefont{Pickett}}, 
  \emph{\bibinfo{title}{arXiv.org:0810.0376 (2008). (To appear in New Journal of Physics)}}

\bibitem[{\citenamefont{lebegue}()\citenamefont{Ma, LU, and Xiang}}]{reznik}
  \bibinfo{author}\bibnamefont{D. Reznik},
  \bibinfo{author}\bibnamefont{K. Lokshin},
  \bibinfo{author}\bibnamefont{D. C. Mitchell},
  \bibinfo{author}\bibnamefont{D. Parshall},
  \bibinfo{author}\bibnamefont{W. Dmowski},
  \bibinfo{author}\bibnamefont{D. Lamago},
  \bibinfo{author}\bibnamefont{R. Heid},
  \bibinfo{author}\bibnamefont{K.-P. Bohnen},
  \bibinfo{author}\bibnamefont{A. S. Sefat},
  \bibinfo{author}\bibnamefont{M. A. McGuire},
  \bibinfo{author}\bibnamefont{B. C. Sales},
  \bibinfo{author}\bibnamefont{D. G. Mandrus},
  \bibinfo{author}\bibnamefont{A. Asubedi},
  \bibinfo{author}\bibnamefont{D. J. Singh},
  \bibinfo{author}\bibnamefont{A. Alatas},
  \bibinfo{author}\bibnamefont{M. H. Upton},
  \bibinfo{author}\bibnamefont{A. H. Said}, 
  \bibinfo{author}\bibnamefont{Yu. Shvyd'ko}, \bibnamefont{and}
  \bibinfo{author}\bibnamefont{T. Egami}, 
  \emph{\bibinfo{title}{arXiv.org:0810.4941 (2008).}}
\end{thebibliography}
\end{document}